\documentclass[pdflatex,sn-mathphys-num]{sn-jnl}% Math and Physical Sciences Numbered Reference Style 

\usepackage{graphicx}%
\usepackage{multirow}%
\usepackage{amsmath,amssymb,amsfonts}%
\usepackage{amsthm}%
\usepackage{mathrsfs}%
\usepackage[title]{appendix}%
\usepackage{xcolor}%
\usepackage{textcomp}%
\usepackage{manyfoot}%
\usepackage{booktabs}%
\usepackage{algorithm}%
\usepackage{algorithmicx}%
\usepackage{algpseudocode}%
\usepackage{listings}%
\usepackage{CJKutf8}

\newcommand{\apjs}{Astrophys. J. Suppl. Ser.}
\newcommand{\apj}{Astrophys. J.}
\newcommand{\mnras}{Mon. Not. R. Astron. Soc.}
\newcommand{\aj}{Astron. J.}
\newcommand{\icarus}{Icarus}
\newcommand{\nat}{Nature}
\newcommand{\apjl}{Astrophys. J. Lett.}
\newcommand{\psj}{Planet. Sci. J.}
\newcommand{\pasp}{Publ. Astron. Soc. Pac.}
\newcommand{\aap}{Astron. Astrophys.}
\newcommand{\pasj}{Publ. Astron. Soc. Jpn.}

\newcommand{\vp}{2012\,VP$_{113}$}
\newcommand{\lele}{Leleākūhonua}

\theoremstyle{thmstyleone}%
\theoremstyle{thmstyletwo}%

\theoremstyle{thmstylethree}%

\begin{document}

\title[A Sedna-like object with $q = 66$ au]{Discovery and Dynamics of a Sedna-like Object with a Perihelion of 66 au}

\author[1]{\fnm{Ying-Tung} \sur{Chen}}\email{ytchen@asiaa.sinica.edu.tw}
\author[2]{\fnm{Patryk Sofia} \sur{Lykawka}}
\author[3]{\fnm{Yukun} \sur{Huang}}
\author[4,5,10]{\fnm{JJ} \sur{Kavelaars}}
\author[4,5,1]{\fnm{Wesley C.} \sur{Fraser}}
\author[6]{\fnm{Michele T.} \sur{Bannister}}
\author*[1]{\fnm{Shiang-Yu} \sur{Wang}}\email{sywang@asiaa.sinica.edu.tw}
\author[1]{\fnm{Chan-Kao} \sur{Chang}}
\author[1,7]{\fnm{Matthew J.} \sur{Lehner}}
\author[8,9]{\fnm{Fumi} \sur{Yoshida}}
\author[10]{\fnm{Brett} \sur{Gladman}}
\author[11]{\fnm{Mike} \sur{Alexandersen}}
\author[1]{\fnm{Edward} \sur{Ashton}}
\author[12]{\fnm{Young-Jun} \sur{Choi}}
\author[1]{\fnm{A. Paula} \sur{Granados Contreras}}
\author[9,3]{\fnm{Takashi} \sur{Ito}}
\author[12]{\fnm{Youngmin} \sur{JeongAhn}}
\author[13]{\fnm{Jianghui} \sur{Ji}}
\author[12]{\fnm{Myung-Jin} \sur{Kim}}
\author[14]{\fnm{Samantha M.} \sur{Lawler}}
\author[15,16]{\fnm{Jian} \sur{Li}}
\author[17]{\fnm{Zhong-Yi} \sur{Lin}}
\author[12]{\fnm{Hong-Kyu} \sur{Moon}}
\author[18,19]{\fnm{Surhud} \sur{More}}
\author[20]{\fnm{Marco} \sur{Muñoz-Gutiérrez}}
\author[21]{\fnm{Keiji} \sur{Ohtsuki}}
\author[5]{\fnm{Lowell} \sur{Peltier}}
\author[11]{\fnm{Rosemary E.} \sur{Pike}}
\author[22]{\fnm{Tsuyoshi} \sur{Terai}}
\author[23]{\fnm{Seitaro} \sur{Urakawa}}
\author[24]{\fnm{Hui} \sur{Zhang}}
\author[13]{\fnm{Haibin} \sur{Zhao}}
\author[16]{\fnm{Ji-Lin} \sur{Zhou}}

% Affiliations
\affil[1]{\orgname{Institute of Astronomy and Astrophysics, Academia Sinica}, \orgaddress{\street{No.1, Sec. 4, Roosevelt Rd}, \city{Taipei}, \postcode{106139}, \country{Taiwan}}}

\affil[2]{\orgname{Kindai University}, \orgaddress{\street{Shinkamikosaka 228-3}, \city{Higashiosaka}, \state{Osaka}, \postcode{577-0813}, \country{Japan}}}

\affil[3]{\orgname{Center for Computational Astrophysics, National Astronomical Observatory of Japan}, \orgaddress{\street{Osawa 2-21-1}, \city{Mitaka}, \state{Tokyo}, \postcode{181-8588}, \country{Japan}}}

\affil[4]{\orgname{Herzberg Astronomy and Astrophysics Research Centre, National Research Council of Canada}, \orgaddress{\street{5071 West Saanich Road}, \city{Victoria}, \state{BC}, \postcode{V9E 2E7}, \country{Canada}}}

\affil[5]{\orgname{Department of Physics and Astronomy, University of Victoria}, \orgaddress{\city{Victoria}, \state{BC}, \postcode{V8W 2Y2}, \country{Canada}}}

\affil[6]{\orgname{School of Physical and Chemical Sciences|Te Kura Mat\={u}, University of Canterbury}, \orgaddress{\street{Private Bag 4800}, \city{Christchurch}, \postcode{8140}, \country{Aotearoa New Zealand}}}

\affil[7]{\orgname{Department of Physics and Astronomy, University of Pennsylvania}, \orgaddress{\street{209 S. 33rd St.}, \city{Philadelphia}, \state{PA}, \postcode{19125}, \country{USA}}}

\affil[8]{\orgname{University of Occupational and Environmental Health}, \orgaddress{\street{1-1 Iseigaoka, Yahata}, \city{Kitakyusyu}, \postcode{807-8555}, \country{Japan}}}

\affil[9]{\orgname{Planetary Exploration Research Center, Chiba Institute of Technology}, \orgaddress{\street{2-17-1 Tsudanuma}, \city{Narashino}, \state{Chiba}, \postcode{275-0016}, \country{Japan}}}

\affil[10]{\orgname{Department of Physics and Astronomy, University of British Columbia}, \orgaddress{\street{6224 Agricultural Road}, \city{Vancouver}, \state{BC}, \postcode{V6T 1Z1}, \country{Canada}}}

\affil[11]{\orgname{Center for Astrophysics \textbar Harvard \& Smithsonian}, \orgaddress{\street{60 Garden Street}, \city{Cambridge}, \state{MA}, \postcode{02138}, \country{USA}}}

\affil[12]{\orgname{Korea Astronomy and Space Science Institute}, \orgaddress{\street{776 Daedeok-daero, Yuseong-gu}, \city{Daejeon}, \postcode{34055}, \country{Republic of Korea}}}

%\affil[13]{\orgname{Center for Computational Astrophysics, National Astronomical Observatory of Japan}, \orgaddress{\street{Osawa 2-21-1}, \city{Mitaka}, \state{Tokyo}, \postcode{181-8588}, \country{Japan}}}

\affil[13]{\orgname{CAS Key Laboratory of Planetary Sciences, Purple Mountain Observatory, Chinese Academy of Sciences}, \orgaddress{\city{Nanjing}, \postcode{210023}, \country{China}}}

\affil[14]{\orgname{Campion College and the Department of Physics, University of Regina}, \orgaddress{\street{3737 Wascana Parkway}, \city{Regina}, \state{SK}, \postcode{S4S 0A2}, \country{Canada}}}

\affil[15]{\orgname{School of Astronomy and Space Science, Nanjing University}, \orgaddress{\street{163 Xianlin Avenue}, \city{Nanjing}, \postcode{210023}, \country{China}}}

\affil[16]{\orgname{Key Laboratory of Modern Astronomy and Astrophysics in Ministry of Education, Nanjing University}, \orgaddress{\city{Nanjing}, \postcode{210023}, \country{China}}}

\affil[17]{\orgname{Institute of Astronomy, National Central University}, \orgaddress{\street{No. 300, Zhongda Rd.}, \city{Zhongli Dist.}, \city{Taoyuan City}, \postcode{32001}, \country{Taiwan}}}

\affil[18]{\orgname{Inter-University Centre for Astronomy and Astrophysics}, \orgaddress{\city{Ganeshkhind}, \city{Pune}, \postcode{411007}, \country{India}}}

\affil[19]{\orgname{Kavli Institute for the Physics and Mathematics of the Universe (WPI)}, \orgaddress{\street{5-1-5 Kashiwanoha}, \postcode{2778583}, \country{Japan}}}

\affil[20]{\orgname{Instituto de Astronomía y Ciencias Planetarias, Universidad de Atacama}, \orgaddress{\street{Copayapu 485}, \city{Copiapó}, \country{Chile}}}

\affil[21]{\orgname{Department of Planetology, Kobe University}, \orgaddress{\city{Kobe}, \postcode{657-8501}, \country{Japan}}}

\affil[22]{\orgname{Subaru Telescope, National Astronomical Observatory of Japan}, \orgaddress{\street{650 North A`ohoku Place}, \city{Hilo}, \state{HI}, \postcode{96720}, \country{USA}}}

\affil[23]{\orgname{Japan Spaceguard Association, Bisei Spaceguard Center}, \orgaddress{\street{1716-3 Okura}, \city{Bisei}, \city{Ibara}, \state{Okayama}, \postcode{714-1411}, \country{Japan}}}

\affil[24]{\orgname{Shanghai Astronomical Observatory, Chinese Academy of Sciences}, \orgaddress{\street{80 Nandan Road}, \city{Shanghai}, \postcode{200030}, \country{China}}}

%\affil[25]{\orgname{CAS Center for Excellence in Comparative Planetology}, \orgaddress{\city{Hefei}, \postcode{230026}, \country{China}}}


\abstract{Trans-Neptunian objects (TNOs) with large perihelion distances ($q > 60$ au) and semi-major axes ($a > 200$ au) provide insights into the early evolution of the solar system and the existence of a hypothetical distant planet. These objects are still rare and their detection is challenging, yet they play a crucial role in constraining models of solar system formation. Here we report the discovery of a Sedna-like TNO, 2023\,KQ$_{14}$, nicknamed `Ammonite', with $q = 66$ au, $a = 252$ au, and inclination $i=11^\circ$. Ammonite's orbit does not align with those of the other Sedna-like objects and fills the previously unexplained `$q$-gap' in the observed distribution of distant solar system objects. Simulations demonstrate that Ammonite is dynamically stable over 4.5 billion years. % with less than 1\% variation in its semi-major axis. 
Our analysis suggests that Ammonite and the other Sedna-like objects may have shared a primordial orbital clustering around 4.2 billion years ago. Furthermore, Ammonite’s stable orbit favors larger orbits ($\sim$ 500 au) rather than closer ones for a large hypothetical planet in present-day trans-Neptunian space. %indicates that a hypothetical planet would possess more distant orbits (a  $\sim$ 500 au), disfavoring closer ones (a  $\sim$ 300-400 au). 
%Our findings provide additional clues for the formation mechanisms of the Sedna-like objects, and new constraints on the hypothetical planet scenario, advancing our understanding of the outer solar system.
}

\maketitle

%%%%%%%%%%%%%%%%%%%%%%%%%%%%%%%%%%%%%%%%%
%%%%%%%%%%%%%%%%%%%%%%%%%%%%%%%%%%%%%%%%%
%%%%%%%%%%%%%%%%%%%%%%%%%%%%%%%%%%%%%%%%%

%\section{Introduction}\label{sec1}
\section{A new Sedna-like object} \label{sec:intro}

The discovery of Sedna \citep{brown04} has initiated an ongoing debate about the formation and evolution of the distant regions of our solar system. Typical trans-Neptunian objects (TNOs) have perihelia $q < 40$ au and their orbits are significantly influenced by Neptune's gravitational perturbations. On the other hand, Sedna-like objects with large semi-major axes ($a > 200$ au) \citep{graham24} and large perihelia ($q > 60$ au) \citep{gomes08} appear to evolve in stable orbits that have remained largely unchanged and not altered by the gravity of Neptune since the formation of the solar system \citep{trujillo20}. No viable transfer mechanisms to raise their perihelia exist with the current configuration of planets.
Their stability suggests that an external gravitational influence beyond those of the currently known solar system planets is required to form their orbits. 

Several scenarios have been proposed to explain the orbits of Sedna-like objects: (1) interactions with a rogue planet-sized body or solar-mass star \citep{morb04, gladman06, portegies15, pfalzner24}; (2) interactions with a hypothetical distant planet \citep{matese05, gomes06, lyk08, matese11, chad14, batygin16}; (3) solar migration within the Milky Way \citep{kaib11}; (4) stellar encounters that took place while the Sun was still a member of its natal star cluster \citep{brasser06, brasser12, kaib08, wajer24}; and (5) the capture of interstellar objects from low-mass stars during the early evolution of the solar system \citep{kenyon04, morb04}. 
To date, only three Sedna-like objects are known, so this population remains poorly understood. However, the discovery of additional objects is particularly challenging due to their great distance from the Sun, resulting in very faint apparent magnitudes.
Increasing the sample of Sedna-like objects is of great interest to better understand the solar system’s history and place stronger constraints on the aforementioned scenarios.

%%%%%%%%%%%%%%%%%%%%%%
Phase two of the ``Formation of the Outer Solar System: an Icy Legacy" project (FOSSIL II) is an extension of the original FOSSIL I project, using the 8.2-m Subaru Telescope on Maunakea (project website: \url{https://www.fossil-survey.org/}).
FOSSIL I utilized a long-term observational cadence repeating on 2-5 pointings to obtain light curves of solar system objects \citep{chang2021,chang2022, ashton23}. 
FOSSIL II surveys $\sim$ 25 deg$^2$ to a limiting magnitude of $m_r\simeq$ 25.2, focusing on (1) the detection of high perihelion TNOs and (2) the dynamical classification and size distribution of resonant TNOs. 
Building on the foundations of shallower surveys like CFEPS (\citet{petit11}, $m_{g}$: 23.5–-24.4), OSSOS (\citet{bannister18}, $m_{r}$: 24.1--25.2) and DES (\citet{pedro22}, $m_{r}$: 23.8), FOSSIL II aims to increase the inventory of small and distant outer Solar System objects, enabling better population modeling and exploration of their properties. 

\begin{figure}[htbp]
    \centering
    \includegraphics[width=1.0\textwidth]{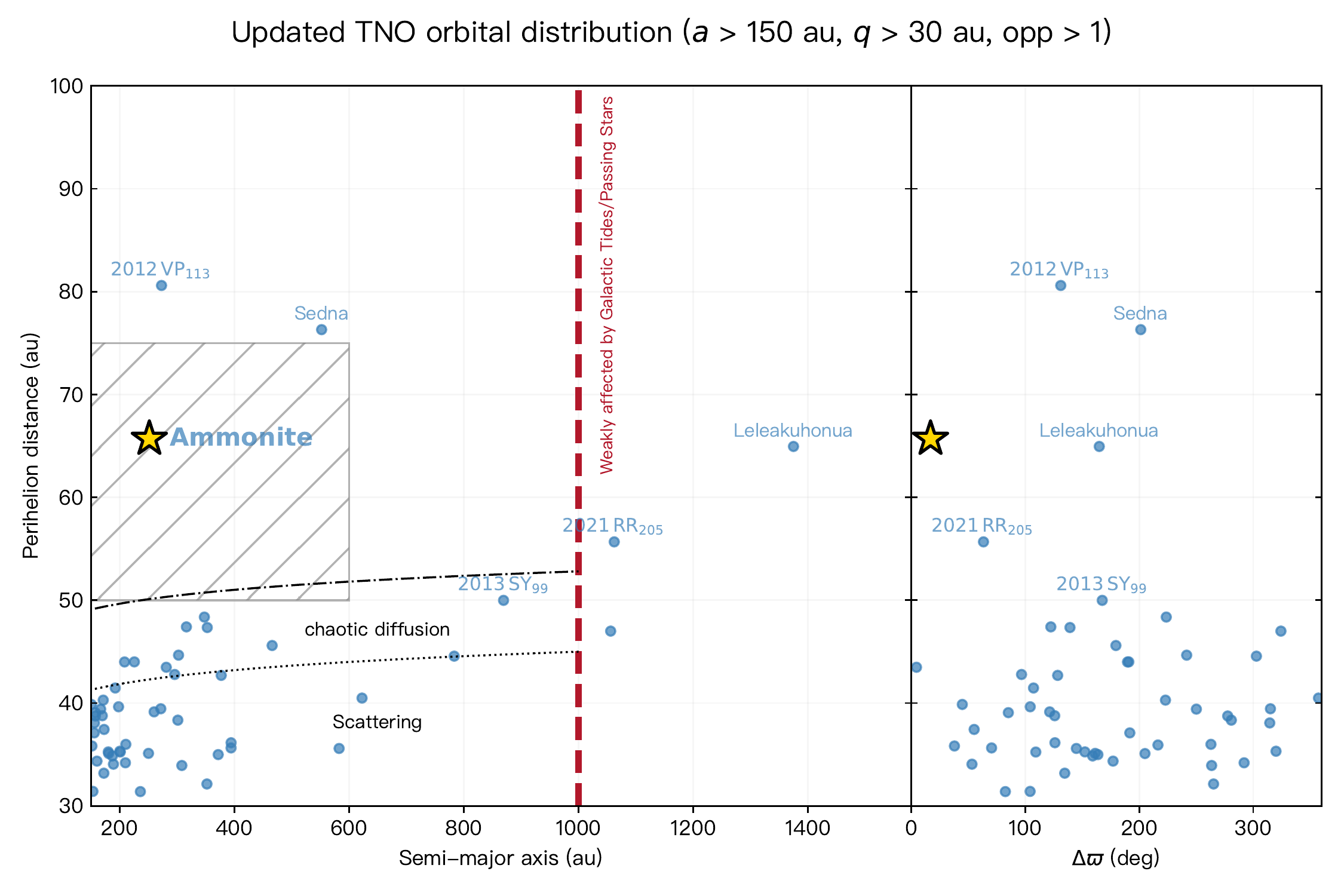}
    \caption{Orbital distribution of known distant TNOs. Objects with observed arcs $\geq~1$ opposition, $a > 150\, \mathrm{au}$ and $q > 30\,  \mathrm{au}$ based on the Minor Planet Center (MPC) database as of February 2025 are plotted. This is an updated version of Fig.~3.1 from \cite{Kavelaars:2020}. The large perihelia objects mentioned in this study are indicated with their names. The location of Ammonite in this plot is marked with a star in both panels. The left panel shows the semi-major axis vs. perihelion distribution, where the dashed vertical red line marks the approximate region where galactic tides and passing stars become significant perturbations on the TNO orbits, while the curved dot-dashed and dotted lines illustrate the upper boundary of chaotic diffusion and gravitational scattering by Neptune, respectively \cite{duncan87, ban17}. The hatched box indicates a region currently lacking any detections, as defined in \cite{Kavelaars:2020}. The right panel presents the distribution of $\Delta\varpi = \varpi-254^\circ$, defined as the difference between the perihelion longitude $\varpi$ ($\varpi = \omega + \Omega$) of each TNO and that of the hypothetical planet proposed in \cite{brown21}; Ammonite falls outside the proposed $\varpi$ clustering of large-$q$ objects. 
    }
    \label{fig:a_q}
\end{figure}

\begin{figure}[h]
    \centering
    \includegraphics[width=0.6\textwidth]{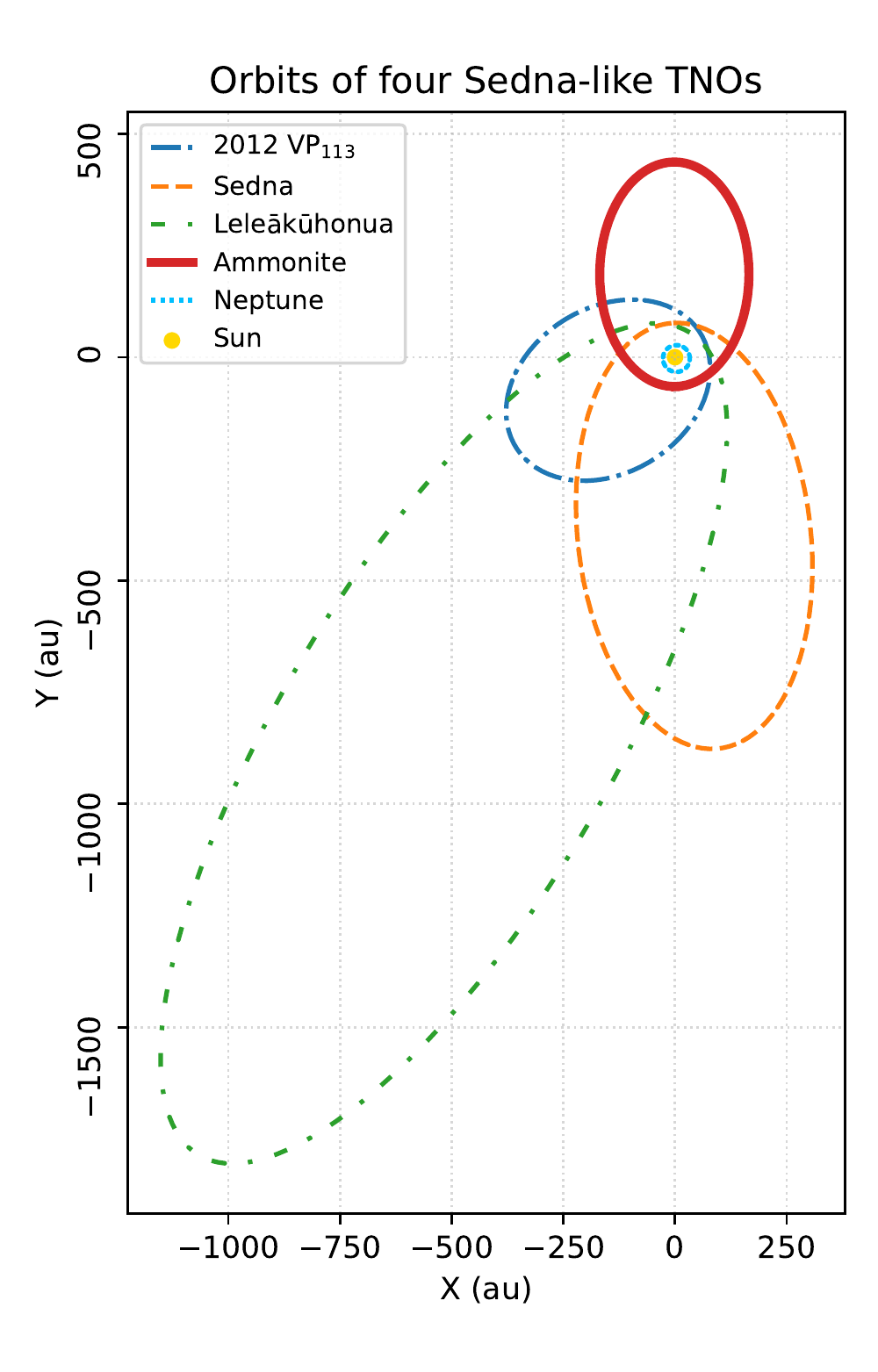}
    \caption{Orbits of the four Sedna-like TNOs projected onto the J2000 ecliptic plane. \hbox{\vp}, Sedna, Leleākūhonua, and Ammonite, with Neptune's orbit around the Sun shown for comparison. The orbital elements used are from the same Minor Planet Center dataset as those in Fig. 1.}
    \label{fig:Hi_q_orbits}
\end{figure}

Within the first year of FOSSIL II observations, we detected an object 2023\,KQ$_{14}$ (nicknamed `Ammonite') with a remarkably high perihelion. % \textcolor{red}{(Note that the MPC designation will be used after we submit its observations, during revision)} 
The barycentric orbit fit, based on a 10.16-year arc (from our observations and archival data, Methods), in the J2000.0 reference frame is: $a = 251.9 \pm 0.3$ au, $e = 0.7383 \pm 0.0003$, $i = 10.98^\circ$ (error $< 0.01^\circ$), $\Omega = 72.104 \pm 0.001^\circ$, $\omega = 198.71 \pm 0.03^\circ$, and $q = 65.9\pm 0.2$ au, with a heliocentric distance of approximately 71.0 au at the time of discovery (2023 May 16.5 UT).
At $q=66$~au, it has the third-largest perihelion among IAU Minor Planet Center-listed objects with a semi-major axis larger than 200 au and multi-opposition observations, following 2012\,VP$_{113}$ ($q=80.6$~au) \citep{chad14} and  (90377) Sedna ($q=76.3$~au) \citep{brown04}, and preceding (541132) Leleākūhonua ($q=65.0$~au) \citep{Sheppard:2019}, 2021\,RR$_{205}$  ($q = 55.6$ au), and 2013\,SY$_{99}$  ($q = 50.0$ au) \citep{ban17}.
The median magnitude of $m_r = 24.6$ corresponds to a diameter of 220-380\,km for albedos of  $p = 0.15-0.05$.
This object fills the ``perihelion gap" of TNO discoveries with $50$~au~$\lesssim q \lesssim$~$75$~au  \citep[see discussion in ][]{Kavelaars:2020}, signaling the importance of distant TNO discoveries to map out the structure of the distant solar system (see Figure~\ref{fig:a_q}).
Meanwhile, Ammonite’s longitude of perihelion is in the opposite direction of the other Sedna-like objects (see Figure~\ref{fig:Hi_q_orbits}).
Its high perihelion suggests the potential for long-term orbital stability, making it valuable for testing the recent hypotheses of primordial clustering of Sedna-like objects \citep{huang24} and the existence of a distant massive planet \citep{lykawka23, brown21, brown24} by analyzing its orbital elements and overall dynamical behavior.

%\begin{figure}[h]
%    \centering
%   \includegraphics[width=0.6\textwidth]{Hi_q_orbits.png}
%    \caption{Orbits of the four largest perihelion TNOs projected onto the J2000 ecliptic plane. \hbox{\vp}, Sedna, Leleākūhonua, and Ammonite, with Neptune's orbit around the Sun shown for comparison. %Both forward and backward integrations were performed using \texttt{Rebound}'s \texttt{WHfast} integrator, with results nearly identical to those obtained with a \texttt{MERCURY} code.
%    }
%    \label{fig:Hi_q_orbits}
%\end{figure}

%%%%%%%%%%%%%%%%%%%%%%%%%%%%%%%%%%%%%%%%%
%%%%%%%%%%%%%%%%%%%%%%%%%%%%%%%%%%%%%%%%%
%%%%%%%%%%%%%%%%%%%%%%%%%%%%%%%%%%%%%%%%%

%\section{Results}\label{sec2}
\section{Results}\label{sec:sim}

%To date, there is no definitive conclusion on the $a$ or $q$ boundary that determines whether an object exhibits Sedna-like characteristics, 
The $a$ and $q$ boundaries that define whether an object exhibits Sedna-like characteristics vary, particularly regarding long-term orbital stability over 4.5 billion years.
Dynamical studies suggest that some objects with similar large-$q$, such as 2013\,SY$_{99}$ and 2021\,RR$_{205}$, may experience gradual orbital migration or diffusion, due to minor perturbations from Neptune and the
influence of galactic tides/passing stars \citep{ban17, huang24}.
Although objects with large semi-major axis and $q> 45$ au tend to remain detached from the influence of the giant planets, their long-term stability needs to be verified through numerical integration \citep{gladman02, lykawka07b, Saillenfest20}. 
A possible way to define the current diffusion boundary analytically is by applying the resonance overlap criterion, as discussed in \citet{batygin21} and \citet{hadden24}, which distinguish between chaotic and non-chaotic regions in $a-q$ space. 

Ammonite’s semi-major axis is close to that of the hypothetical planet proposed by \citet{batygin16} and \citet{lykawka23}, as well as to the boundary of the Neptunian mean motion resonances (MMRs) in which an object may experience perihelion-raising  through von Zeipel-Lidov-Kozai dynamics or MMRs \citep{zeipel1910, gallardo12, ito19, graham24, lykawka07}. Additionally, Ammonite provides a good test case for the primordial clustering of Sedna-like objects generated by a transient planetary body.
Therefore, we performed the following numerical simulations to verify the orbital evolution of Ammonite.

%%%%%%%%%%%%%%%%%%%%%%%%%%%%%%%%%%%%%%%%%
%%%%%%%%%%%%%%%%%%%%%%%%%%%%%%%%%%%%%%%%%
%%%%%%%%%%%%%%%%%%%%%%%%%%%%%%%%%%%%%%%%%

\subsection{Long-term stability}\label{sec:stabilty}

The results of both forward and backward simulations using two independent codes indicate similar stability (Methods), with mean variations of the semi-major axis and eccentricity remaining under 1\% for the best fit $a$, possible highest $a$ and possible lowest $a$ orbits, as shown in Supplementary Figure 1. The inclination of the clones of Ammonite oscillates between 8 and 11 degrees throughout the simulations. 
%None of the orbital elements show signs of resonance trapping or sticking behavior.
%None of the orbital elements of the remaining clones exhibit significant or consistent signs of resonance trapping or sticking behavior.
%The three representative clones and the remaining clones show no significant or consistent signs of resonance trapping or sticking behavior in their orbital elements.
%None of 1000 clones show no significant signs of closest 24:1 resonance behavior in their orbital elements.
None of the 1000 clones show evidence of resonant behavior in their orbital history; the closest major resonance (the 24:1 at a=250.1 au) is about 5-sigma away from the best fit.
This orbital evolution is consistent with studies of this region of orbital parameter space, which indicate that objects with $a > 200$ au and $q>60 $ au experience minimal orbital evolution in the timescale of $10^9$ years \citep{lykawka07, gallardo12}. %, although they may exhibit chaotic behavior over longer periods \citep{saillenfest19}.
The orbital evolutions clearly demonstrate this object's similarity to other Sedna-like objects. Based on this, we conclude that it can be identified as the fourth Sedna-like object discovered to date, with the third-largest perihelion.

TNO discoveries to date suggest the presence of a ``perihelion gap": an apparent underpopulated region among the orbital parameters of TNOs with $150 < a\lesssim 600$~au and perihelia between roughly 50--75~au \citep{Sheppard:2019,Kavelaars:2020,Oldroyd:2021}.
Crucially, the gap is not generated by the limitation of sensitivities of the surveys as more distant discoveries do occur (see Figure~\ref{fig:a_q}).
If this gap does exist, it could be considered as a structural feature of the population, with implications for distinguishing between orbital formation and evolutionary mechanisms, such as the semi-major axis diffusion seen in larger-$a$ orbits.
Ammonite is the first TNO with $150 <a<600$~au to have a perihelion in this gap.
As demonstrated in the Supplementary Figure 1, Ammonite’s $a$ is sufficiently small that it remains stable so that diffusion does not explain its orbit  \citep{ban17, Kavelaars:2020}.
%On the other hand, since Ammonite has a larger $q$ than that of Leleākūhonua, its discovery does not change the somewhat narrower perihelion gap proposed by \citet{Oldroyd:2021}, though Leleākūhonua’s large $a$ makes its $q$ susceptible to galactic tide influences. 
%Consistent with the gap's sensitivity expectations, at $m_r = 24.6$, 
%Ammonite is distinctly brighter than FOSSIL II's limiting design sensitivity of $m_r = 25.2$ or in-night magnitude of $m_r = 25.0$.
%We defer a full assessment of the FOSSIL II detection sensitivity to the release of the complete TNO sample.
%This detection suggests that the perihelion gap may still exists in the range $50\lesssim q\lesssim65$~au, 
Therefore, a formation mechanism is still undecided but is required to populate orbits throughout the perihelia range. Future surveys with more detections are necessary to determine if there is any distribution gap associated with the population.

\subsection{Exploring a Possible Primordial Orbital Alignment}\label{sec:cluster}

Recently, \citet{huang24} examined the orbital histories of the three previously known Sedna-like objects: Sedna, \hbox{\vp}, and \hbox{\lele}. 
That study revealed an intriguing result where their longitudes of perihelion ($\varpi$) converged to a narrow cluster around 200 degrees 4.5~Gyr ago.
Although this clustering hints at a primordial event that elevated their perihelia, additional discoveries and analyses are needed to solidify this picture.
Here, we extend the analysis by including Ammonite and performing a similar backward integration using the same parameters as in \citet{huang24}.

\begin{figure}[ht]
    \centering
    \includegraphics[width=\textwidth]{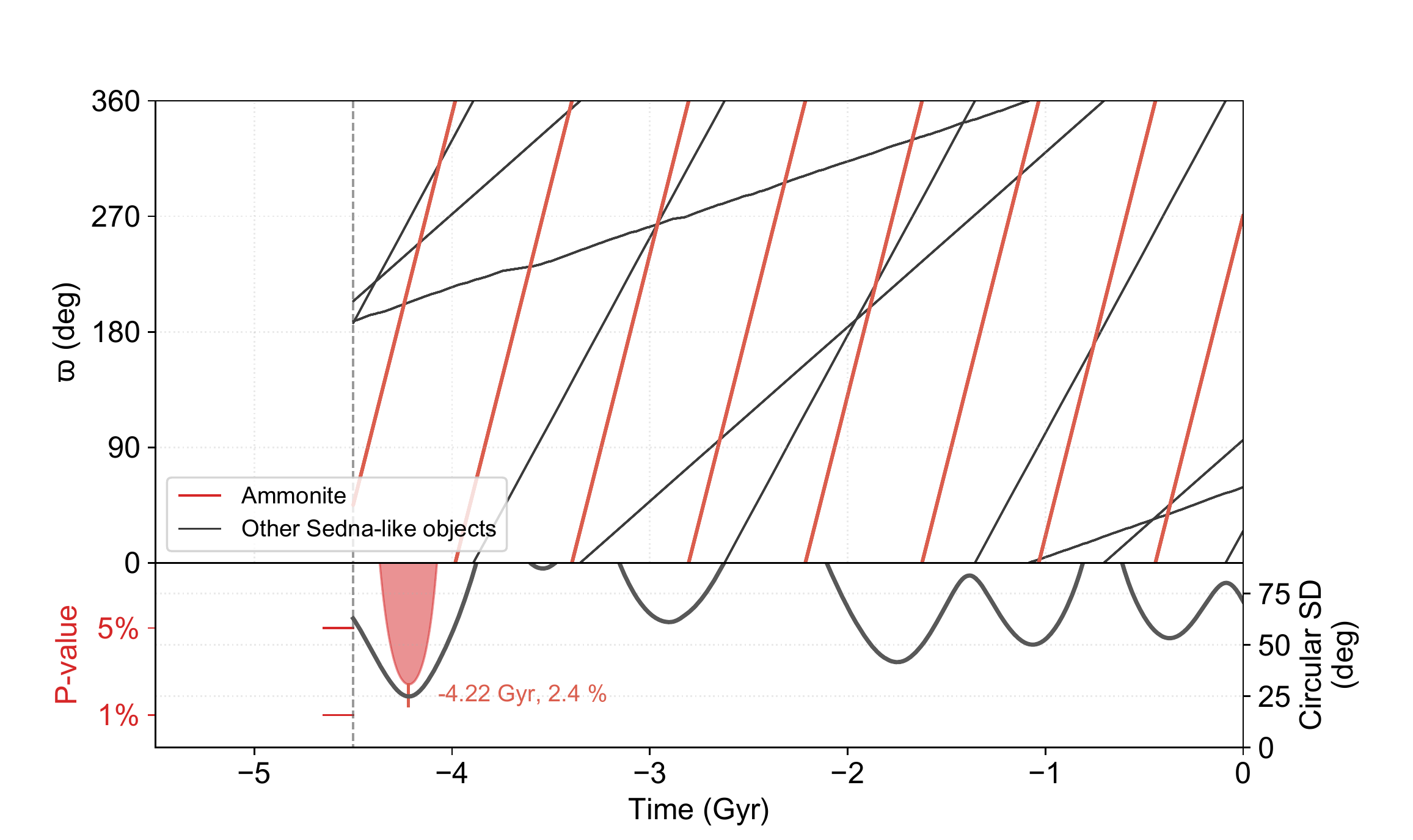}
    %\caption{The simulation of Ammonite suggests a late primordial clustering around -4.2 Gyr, approximately 300 Myr post the solar system's formation, with over 97\% confidence.}
    \caption{Time evolution and statistical analysis of the perihelion longitudes of four Sedna-like objects. Upper: past evolutions of longitudes of perihelion ($\varpi$) for Ammonite (red) and the other three Sedna-like objects (black). Lower: the circular standard deviation (SD) of the four angles (black) and the statistical confidence ($p$-value, red shaded) that they are generated from a uniform distribution. The addition of Ammonite suggests a late primordial clustering around $4.2$~Gyr ago compared to \citep{huang24}, approximately 300 Myr after the solar system's formation, with over 97\% confidence.} 
    \label{fig:cluster}
\end{figure}

Our results indicate a comparable clustering event around 4.2~Gyr ago, roughly 300~Myr after the solar system’s formation, with a confidence level (measured by the Rayleigh test of uniformity) exceeding 97\% (see Figure~\ref{fig:cluster}). 
% We calculate the $p$-value using a Rayleigh test of uniformity \cite{huang24}, which indicates less than a 5\% probability that the observed clustering is due to random chance. 
In addition, we performed 10,000 Monte Carlo simulations to test the robustness of this early clustering. Our results reveal that fewer than 7.7\% of randomly shifted orbital histories exhibit a stronger $\varpi$ clustering (between 4.55 and 4.16~Gyr ago, as detailed in the Methods) than what we observe among the four Sedna-like objects.

However, we note that this level of significance corresponds to slightly less than 2$\sigma$, and the inclusion of Ammonite results in a somewhat looser and delayed clustering relative to the findings of \citet{huang24}. 
If future observations confirm a more pronounced and statistically robust clustering, it could imply that a transient planetary perturber \citep[e.g.][]{gladman06,Huang.2023t} played a role early in the solar system’s history. Following such an event, the clustering might have gradually dispersed due to the differential precession of the apsidal lines driven by the four giant planets. Stellar flyby models \citep[e.g.][]{morb04, brasser12, wajer24}, on the other hand, do not produce a clustered $\varpi$ \citep{huang24}. Further observations, particularly a more precise refinement of \hbox{\lele}'s orbit and the discovery of new Sedna-like TNOs, will be essential to confirm or challenge this tentative primordial alignment and to better constrain the formation history of the early solar system.

\subsection{Interaction with a hypothetical planet}\label{sec:p9}

It is important to note that Ammonite’s $\varpi$ and $\Omega$ do not align or cluster with those of Sedna, \hbox{\vp}, and \hbox{\lele} (see right panel of Figure~\ref{fig:a_q} and Figure~\ref{fig:Hi_q_orbits}). %This distinction is crucial in understanding the potential dynamical mechanisms operating in the distant solar system.
A present-day planet has been proposed as a mechanism for gravitationally influencing and clustering the orbits of distant TNOs \citep{batygin16}.
If this massive body indeed exists in this region, the stability of Sedna-like objects could serve as a test.
In other words, the presence of Sedna, \hbox{\vp}, \hbox{\lele}, and Ammonite should indicate either negligible or strong dynamical interactions with the putative planet. %which can be assessed by the stability of their orbits over long timescales, 
Therefore, we employed the \texttt{MERCURY} integrator, using the same clone generation as for the orbital stability analysis (Methods), to simulate three clones of each of the four Sedna-like objects for 1 Gyr, applying an $a>$ 10,000\,au criterion for ejection.

We incorporate planetary orbits proposed in previous studies into these simulations to investigate the hypothetical planet's influence on the stability of the four TNOs.
\citet{brown21} estimated the mass and orbit for the hypothetical planet of: $M = 6.2^{+2.2}_{-1.3} \, M_{\oplus}$, $a = 380^{+140}_{-80}$ au, $i = 16^\circ \pm 5^\circ$, and $q = 300^{+85}_{-60}$ au.
We selected four sets of orbital elements of the hypothetical planet in this investigation: (a) maximum likelihood, (b) maximum perihelion distance, (c) minimum perihelion distance from Table 2 of \citet{brown21}, and (d) nominal values from Figure 8 of \citet{brown21} (see Table~\ref{tab:hyp_planet_params}).
Because the mean anomaly of the hypothetical planet is not well constrained in previous studies, we selected mean anomaly values of 0$^\circ$, 60$^\circ$, 120$^\circ$, 180$^\circ$, 240$^\circ$, and 300$^\circ$ in our simulations.
As shown in Table~\ref{tab:hyp_planet_params}, the survival rates of Sedna, \hbox{\vp}, and \hbox{\lele} were relatively high. Only 4 of 216 clones (one Sedna clone and three \hbox{\vp} clones) were ejected before the end of the 1 Gyr simulation. In contrast, most Ammonite clones (47 out of 54), except those in case (b), experienced orbital instability and were ejected in the simulation. This is expected, as Ammonite's nominal orbit has a higher probability of close encounters with the hypothetical planet of a similar orbit.

\begin{figure}[ht]
    \centering
    \includegraphics[width=0.8\textwidth]{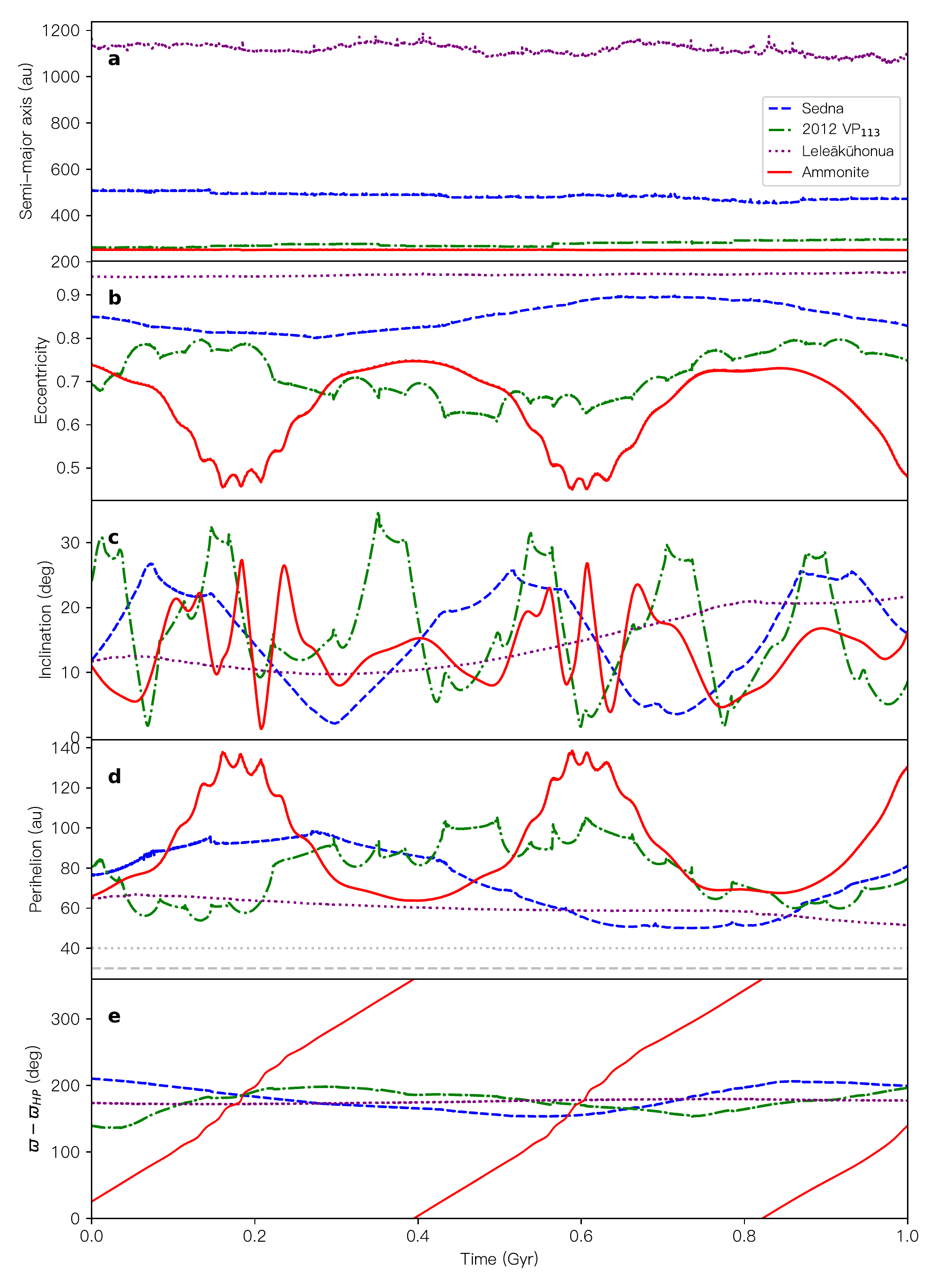}
    \caption{Orbital stability of Sedna-like objects under the influence of a hypothetical planet. The stability of four Sedna-like objects using the nominal hypothetical planet parameters from \citet{brown24} ($M = 6.6\,M_{\oplus}$, $a = 500$ au, $Q=630$ au ($e=0.26$), $i= 15.6^\circ$). 
    %Panels (a)–(e) show the time evolution of $a$, $e$, $i$, $q$, and $\varpi$, respectively, for all four objects. 
    Panels (a)–(e) show the time evolution of $a$, $e$, $i$, $q$, and $\Delta\varpi = \varpi - \varpi_{\scriptscriptstyle HP}$ (the difference between each object’s longitude of perihelion and that of the hypothetical planet), for all four objects, respectively.
    The results suggest that Sedna, 2012\,VP$_{113}$, and Leleākūhonua are strongly influenced and clustered in longitude of perihelion ($\varpi$) with respect to this hypothetical planet, whereas Ammonite behaves differently. In the perihelion evolution panel, the dotted and dashed lines represent $q=40$ au and $q=30$ au, respectively.}
    \label{fig:ammonite_HP}
\end{figure}

\citet{brown24} later updated their estimates of planet's orbital properties to $a$ = $500^{+170}_{-120}$ au, $M$ = $6.6^{+2.6}_{-1.1} M_{\oplus}$, and aphelion distance of $630^{+290}_{-170}$ au.
We also performed simulations with the updated nominal orbit (Table~\ref{tab:hyp_planet_params}, case (e)), resulting in the ejection of only one \hbox{\vp} clone by the end of 1 Gyr.
The nominal orbit elements used here are similar to those in case (b), implying that a planet’s orbit with larger $a$ and $q$ has a lower likelihood of close encounters with Ammonite.
Figure~\ref{fig:ammonite_HP} illustrates the stability of nominal orbits of the four Sedna-like objects from a representative simulation.
The simulation results indicate that Sedna, \hbox{\vp}, and \hbox{\lele} experience strong gravitational interactions with the hypothetical planet, consistent with \citet{brown21}, and exhibit significant clustering of their $\varpi$.
This suggests that these Sedna-like objects would be gravitationally shepherded by such a planet, maintaining relatively stable configurations over the 1 Gyr.

In contrast, Ammonite shows quite different behavior with the \citet{brown24} planet compared with the other three Senda-like objects. Some Ammonite clones with different mean anomaly values exhibited only temporary clustering, suggesting slight variations in initial orbital parameters could affect the simulation results. 
However, the different stability of Ammonite compared to the other three objects suggests that the definition of Sedna-like objects should consider sub-dynamical populations if this hypothetical planet exists.
It is worth noting that the semimajor axis of Ammonite (252 au) lies near the transition ``wall" suggested by \citet{brown21} between the nearby uniformly-$\Delta\varpi$-distributed population and the distant clustered population.
This proximity to the transition region may explain why some clones of Ammonite's orbit still experience temporary clustering due to the gravitational influence of the hypothetical planet. %is less perturbed and less prone to clustering, as such behavior could be expected under the specific planetary parameters considered here.
Additionally, the orbital pole positions of the four Sedna-like objects show a generally random distribution, rather than the significant clustering of pole positions seen in the sample of objects with $150 < a < 1000$ au and $q > 42$ au, as illustrated in Figure 2 of \citet{brown21}.
The different orbital influence by the hypothetical planet on Ammonite provides a valuable contrast, emphasizing the range of dynamical behaviors that Sedna-like objects may exhibit in response to the presence of this hypothetical planet.
%It is important to highlight that primordial alignment and the presence of a distant planet today are mutually exclusive \citep{huang24}. 
It is important to highlight that primordial alignment (calculated through the perturbations of the four giant planets) and the current presence of a distant planet are mutually exclusive.
%If the alignment is primordial, there is no current influence from a distant planet; if a distant planet exists today, the alignment cannot be purely primordial. 
Further discoveries of Sedna-like objects will clarify which external gravitational influence raised the perihelion of these objects.

%\section{Discussions}
\section{Conclusion}\label{con}
%The discovery of ``Ammonite", an object on a distant Sedna-like orbit with a perihelion of $\sim66$ au, offers a valuable opportunity to evaluate current models of outer solar system formation and evolution. The confirmed stable orbit of Ammonite through simulations constrains the possible orbital parameter range of a hypothesized distant and currently undetected planet. 
The discovery of ``Ammonite", the first anti-cluster Sedna-like object with the third largest $q$ among all TNOs, offers a valuable opportunity to evaluate current models of outer solar system formation and evolution. 
With a perihelion of $\sim66$ au, Ammonite's confirmed stable orbit through simulations provides constraints on the possible orbital parameter range of a hypothesized distant and currently undetected planet.
Meanwhile, simulations including all four Sedna-like object shows they may have experienced a primordial clustering of perihelion longitudes around 4.2 Gyr ago.
%With an extended observational arc of over ten years, Ammonite's secure orbit contributes to the study of distant TNOs and their dynamics.

%Simulations indicate that Ammonite's orbit is stable over $\pm$4.5 Gyr, with less than 1\% mean variations of $a$.
%Ammonite potentially provides evidence that the formation mechanism of high-$q$ objects has to leave a gap when populating perihelia throughout the parameter range, above where diffusion can make orbits.\textcolor{red}{(I do not understand "what make orbits" means)}
%Ammonite potentially provides evidence that the formation mechanism of high-$q$ objects likely leaves a narrower gap when populating perihelia across the parameter range, above which diffusion dominates the mechanism that forms medium-$q$ objects.
%Ammonite's orbit fills a previously observed $q$-gap, potentially indicating that the gap may only appear in smaller samples or represent a narrower distribution when populating perihelia across the parameter range.
%Additional simulations suggest that Ammonite, like the other three largest-$q$ TNOs, may have experienced a primordial clustering of perihelion longitudes around 4.2 Gyr ago.
%Further simulations involving Sedna, \hbox{\vp}, \hbox{\lele} under the influence of a hypothetical planet show that Ammonite's orbit is not stable if the planet has $a\sim380$ au. However, this result is much less constrained for a planet with $a\sim500$ au, compared to the other three largest-$q$ objects. 
These findings highlight the diversity of orbital properties and dynamical behaviors among distant solar system objects. 
Future large surveys will be the key to increasing the number of large-$q$ objects and refining our understanding of the dynamical processes shaping the outer solar system.

\section{Methods} \label{sec:medthod}
\subsection{Observations and orbit fit}

The FOSSIL II survey is designed for the pre-discovery (known as precovery), discovery, and recovery of TNOs, using the Hyper Suprime-Cam (HSC) \citep{miyasaki18} on Subaru Telescope on Maunakea. 
With 16 closely-packed pointings of HSC each imaged with 270s exposures, FOSSIL II covers 25 deg$^2$ to an expected limiting magnitude of $m_{r} \simeq 25.6$. 
The observation plan was structured into three epochs, each separated by an interval between 60 and 90 days, with three exposures per night spaced by $\sim80$ minutes (a `triplet') to confirm the on-sky motion of solar system objects. 
The initial orbit determination required at least two triplets separated by a few days. The time allocation included three half-nights for precovery and discovery in March and May 2023, and an additional two half-nights for recovery observations in June, with the goal of obtaining at least one triplet in each month.
Precovery observations were successfully carried out using the new `EB-gri' broad-band filter \citep{fraser24} with 270s exposures from March 18-20, 2023, over three half-nights (acquiring three triplets). 
However, due to weather, the discovery observations in May were limited to only one triplet on May 16, 2023, achieving an average limiting magnitude of $m_{r} \simeq 25.2$.
Unfortunately, all the recovery time in June 2023 was lost due to technical issues.
Additional recovery observations in $r$-band with longer exposures at 380s were awarded for three quarter-nights of observation in August 2023. 

In the preliminary FOSSIL II TNO candidate list, we  identified Ammonite as an object with an extraordinarily high perihelion and barycentric distance. 
However, due to the slow movement of TNOs, the five-month observation arc in 2023 was insufficient to accurately determine the perihelion distance. 
Therefore, a CFHT DDT proposal was submitted for two triplets of observations in July 2024. 
CFHT secured triplets on two different nights with 380s exposures in $w$-band, which improve the orbit determination of Ammonite.
We then explored archived data through the Solar System Object Image Search (SSOIS; \url{https://www.cadc-ccda.hia-iha.nrc-cnrc.gc.ca/en/ssois/}) \citep{Gwyn:2012}. 
The point spreading functions of the moving object %(referred to as "counterparts") 
were identified within a 1-sigma error prediction ellipse after incorporating the CFHT DDT measurements. 
We note that one of the precoveries in the 2021 DECam archive was identified in the raw (unreduced) image, because the pixels within the prediction ellipse in the publicly-calibrated image were resampled due to a bright star streak.
After adding the 2021 measurements, even earlier DECam precoveries were identifiable in 2014, near the center of the 1-sigma error ellipse. Overall, these result in a total arc-length of 10.16 years for Ammonite.
Supplementary Table 1 summarizes the FOSSIL II and CFHT observations as well as the archival data we used in this study.

The barycentric orbit fit, based on \citet{ber00} (see \url{https://web.sas.upenn.edu/garyb/software/}), to the observations (Supplementary Table 1) in the J2000.0 celestial reference frame is as follows: semi-major axis $a = 251.9 \pm 0.3$ au, eccentricity $e = 0.7383 \pm 0.0003$, inclination $i = 10.98$ degrees (with error $< 0.01^\circ$),  ascending node $\Omega = 72.104 \pm 0.001$ degrees, argument of periapsis $\omega = 198.71 \pm 0.03$ degrees, and perihelion distance $q = 65.9\pm 0.2$ au, determined from a 10.16-year arc with mean residual of 0.1\,$^{\prime\prime}$. 

\subsection{Long-term stability}

As chaotic diffusion \citep{saillenfest19} and minor perturbations may behave differently with various integrators, we performed N-body simulations using two codes: \texttt{hybrid symplectic/Bulirsch-Stoer} in \texttt{MERCURY} \citep{chambers99} and \texttt{WHFast} in \texttt{REBOUND} \citep{rein12}.
The simulations were conducted with a time step of 180 days, integrating the orbit both forward and backward over 4.5\,Gyr. To improve computational efficiency, 
the mass of the terrestrial planets was incorporated into the Sun, leaving only the four giant planets as massive perturbers in our nominal simulations.
In addition to the nominal orbit, we selected another two clones with the highest and lowest values of $a$ from a set of 1000 clones generated using a covariance matrix and a Gaussian random distribution within 3-sigma of the orbital element errors to accurately account for orbital uncertainties, following the classification method of \citet{gladman08}. 

% Methods about premordial alignment
\subsection{Compatibility with a primordial orbital alignment}

Based on different hypotheses for the formation of sedna-like objects, the timing when they were primordially implanted to their current orbits varies. For the stellar encounter model, \citet{nesvorny23b} argue that the sedna-like objects were implanted $\sim$10 Myr after the gas disk was dispersed, whereas a rogue planet model generally requires $\sim$100--300 Myr \citep{gladman06, Huang.2023t} for the planet to continuously lift planetesimals out of the primordial scattering disk. Assuming the rogue planet was initial formed in the giant planet region, it was most likely scattered to large-$a$ orbit right after the giant planet instability \citep{nesvorny12, griveaud24}, which also triggered Neptune's migration into the outer planetesimal disk and supplied icy bodies to be implanted into the region of sedna-like objects. Previous studies \citep{nesvorny18b, avdellidou24, edwards24} showed that the instability must have occurred within 100 Myr after the dispersal of the gas disk. Therefore, in a rogue planet hypothesis, the timing of implantation (which corresponds to the timing when the postulate primordial clustering is tightest) varies from $\sim$100 Myr to $\sim$400 Myr. 

Therefore, we argue that $t = 10$ -- 400~Myr after the disk gas dispersal should be treated as the potential “time of interest” for the primordial alignment. This time interval corresponds to -4.55 Gyr to -4.16 Gyr if one assumes the gas disk disperse 10 Myr after the formation of the Solar System 4.57 Gyr ago. We thus evaluated the significance of the clustering by assuming random current values of $\varpi$ for each object, while accounting for the precession rates of the four Sedna-like objects: $610^\circ$/Gyr for Ammonite, $136^\circ$/Gyr for Sedna, $284^\circ$/Gyr for \hbox{\vp}, and $51^\circ$/Gyr for \hbox{\lele}. 
Notably, these precession rates reflect the gradual evolution of their orbits due to the perturbations from the giant planets.
However, it is important to recognize that this analysis does not account for the orbital uncertainties of these objects, which could affect the precise timing and extent of the clustering.

\backmatter

%\bmhead{Supplementary information}

%If your article has accompanying supplementary file/s please state so here. 
%Authors reporting data from electrophoretic gels and blots should supply the full unprocessed scans for key as part of their Supplementary information. This may be requested by the editorial team/s if it is missing.
%Please refer to Journal-level guidance for any specific requirements.

\bmhead{Data availability}
The observational data used in this study will be deposited at the Minor Planet Center (\url{https://www.minorplanetcenter.net/}) upon publication.

\bmhead{Code availability}
The \texttt{MERCURY} N-body integrator (version 6) is available from a public GitHub mirror at \url{https://github.com/smirik/mercury}. The \texttt{REBOUND} N-body code can be accessed at \url{https://github.com/hannorein/rebound}.
%\texttt{MERCURY} and \texttt{REBOUND} are open-access codes. 
The relevant parameters are provided with this paper. 
%Orbfit, REBOUND and Mercury are open access codes.

\bmhead{Correspondence and requests for materials}
Correspondence and requests for materials should be addressed to Y.-T.C.

\bmhead{Acknowledgements}
%\textcolor{red}{need to be finished}
This research is based in part on data collected at the Subaru Telescope, which is operated by the National Astronomical Observatory of Japan (NAOJ). We are honored and grateful for the opportunity of observing the Universe from Maunakea, which has the cultural, historical, and natural significance in Hawaii.
Based on observations obtained with MegaPrime/MegaCam, a joint project of CFHT and CEA/DAPNIA, at the Canada-France-Hawaii Telescope (CFHT) which is operated by the National Research Council (NRC) of Canada, the Institut National des Science de l'Univers of the Centre National de la Recherche Scientifique (CNRS) of France, and the University of Hawaii. The observations at the Canada-France-Hawaii Telescope were performed with care and respect from the summit of Maunakea which is a significant cultural and historic site. The numerical orbit integration carried out for this study was partially performed at Center for Computational Astrophysics (CfCA), NAOJ
S.-Y.W. acknowledges support from National Science and Technology Council of Taiwan with grant No. 113-2112-M-001-003-. K.O., F.Y., and T.T. acknowledge support from JSPS Grants-in-Aid for Scientific Research JP23K22557. Y.H. acknowledges support from JSPS Grants-in-Aid for Early-Career Scientists No. 25K17460. J.J. acknowledges support from the National Natural Science Foundation of China (Grant Nos. 12150009,12033010)

\bmhead{Author Contributions}
Y.-T.C. analyzed the data, generated the candidate list, proposed follow-up observations, conducted numerical simulations, and drafted the paper. 
P.S.L. and Y.H. contributed to numerical simulations, helped analyze the long-term stability and orbital evolution, and prepare the simulation part of the paper. 
J.J.K. planned and analyzed the CFHT DDT follow-up observations, confirming the uncertainty of the perihelion. 
W.C.F., M.T.B., and S.-Y.W. helped with the DDT proposal and contributed to dynamical discussions. 
M.T.B. and S.-Y.W. assisted in finalizing the paper. 
S.-Y.W. and C.-K.C. analyzed the data, and validated the candidates.
S.-Y.W. and F.Y. organized the FOSSIL II Subaru proposals and F.Y. served as the principal investigator of the project. 
S.M.L., R.E.P., J.J, J.J.K., and B.G. also contributed to discussions on long-term stability and orbital evolution.
T.I., F.Y., T.T., Y.-T.C., and C.-K.C. performed FOSSIL II observations
Y.-T.C., P.S.L., J.J.K., W.C.F., S.-Y.W., C.-K.C., M.J.L., F.Y., M.A., E.A., Y.-J.C., A.P.G.C., T.I., Y.J., J.J., M.-J.K., S.M.L., J.L., Z.-Y.L., H.-K.M., S.M., M.M.-G., K.O., L.P, R.E.P., T.T., S.U., H.Z. (Hui), H.Z. (Haibin), and J.-L.Z. all helped on the FOSSIL proposal and planned the observations.
All authors were given the opportunity to review the results and comment on the manuscript.

%Acknowledgements are not compulsory. Where included they should be brief. Grant or contribution numbers may be acknowledged.

%Please refer to Journal-level guidance for any specific requirements.

\bmhead{Competing interests}
The authors declare no competing interests.

\clearpage

\section*{Tables}

\begin{table*}[ht]
\centering
\caption{The parameters of hypothetical planet and simulation results \label{tab:hyp_planet_params}}
\begin{tabular*}{\textwidth}{@{\extracolsep\fill}lccccccc}
\toprule
  & Mass ($M_{\oplus}$) & $a$ (AU) & $i$ ($^\circ$) & $e$ & $\varpi$ ($^\circ$) & $\Omega$ ($^\circ$) & Ejected Clones$^{\dagger}$ \\
\midrule
(a) & 5   & 300    & 17    & 0.15  & 254    & 108   & 15, 0, 0, 0 \\
(b) & 7   & 500    & 20    & 0.15  & 256    & 94    & 0, 0, 1, 0 \\
(c) & 6   & 310    & 15    & 0.20  & 252    & 92    & 17, 0, 2, 0 \\
(d) & 6.2 & 382.4  & 15.6  & 0.20  & 247.6  & 97.5  & 15, 1, 0, 0 \\
(e) & 6.6 & 500    & 15.6  & 0.26  & 247.6  & 97.5  & 0, 0, 1, 0 \\
\botrule
\end{tabular*}
\begin{tablenotes}
\item $^{\dagger}$ The numbers in ``Ejected Clones" are for Ammonite, Sedna, \vp, and \lele, respectively. Each TNO has 18 initial clones.
\item (a) \citet{brown21} max likelihood. 
\item (b) \citet{brown21} max $q$.
\item (c) \citet{brown21} min $q$ \& $i$.
\item (d) \citet{brown21} peak of contour.
\item (e) \citet{brown24} parameters.
\end{tablenotes}
\end{table*}

\clearpage

\section*{Figure Legends/Captions}

\clearpage


\end{document}